# SensoPatch: A reconfigurable haptic feedback with high-density tactile sensing glove

Yanisa Angkanapiwat[1,2], Ariel Slepyan[2], Rebecca J. Greene[2], Nitish Thakor[1,2]

[1]Department of Biomedical Engineering, Johns Hopkins University, Baltimore, MD, USA
[2]Department of Electrical and Computer Engineering, Johns Hopkins University, Baltimore, MD, USA

*Abstract*—Haptic feedback is integral to the improved experience of prosthetic users and the reduction in prosthesis rejection. Unfortunately, existing haptic feedback systems often face challenges that impede their transition to clinical use including lack of intuitive control, reproducibility, and personalized customization. Prior studies have explored various methods to encode tactile information and deliver vibration feedback. However, a comprehensive study comparing performance across different stimulation locations and feedback modalities for wearable devices is absent and there is no robust test platform. This paper proposes an open-source reconfigurable haptic feedback system which features 25 sensors and wireless communication to allow customized number of vibration motors, adjustable motor placement, and programmable encoding of tactile data to change feedback modalities. To demonstrate the potential studies that can be investigated using the proposed system, we conducted two preliminary experiments: 1) to assess the vibration discrimination accuracy on three body parts (upper arm, shoulder, and lower back)  2) to evaluate the effect of six methods of mapping tactile data to varying number of motors on object manipulation. This proposed tool utilizes low-cost off-the-shelf components, enabling large-scale comparative studies of feedback modalities and stimulation sites to optimize vibrotactile feedback and facilitate its deployment in upper limb prostheses.

## I. Introduction

Haptic feedback plays a pivotal role in enhancing the sensory experience for amputees and is related to increased prosthesis functionality, higher sense of embodiment, and improved object manipulation [1, 2]. Vibrotactile stimulation is one of the most common non-invasive feedback methods due to its small size, low cost, and low power consumption [3]. Previous research on wearable devices for upper limb amputees has investigated several locations for stimulus application such as upper arm, forearm, fingertip, foot, waist, and neck [4]. Additionally, prior studies have explored several feedback modalities to represent tactile information from different parts of the hand using different number of vibration motors. For example, Nabeel et al. [5] examined the use of 2 vibration motors on forearm for delivering proportional feedback corresponding to force on one finger. Li et al. [6] designed a system with 5 vibration motors which delivers feedback from sensations on each finger to the back of the the corresponding finger of the opposite hand. Another system proposed by Aug et al. [7] consists of 6 vibration motors on upper arm for conveying tactile information from all fingers and palm.

Despite the abundance of system designs, few studies have directly compared the performance of different feedback strategies and stimulation sites [8]. Although adjustability of location is one of the prosthesis user priorities for feedback [9], existing wearable haptic feedback devices are typically designed for only one body location and intended to communicate sensations from specific parts of the hand, which restricts their adaptability and renders comparison between studies challenging [10]. The absence of a reconfigurable haptic feedback system poses a significant barrier to optimizing feedback delivery and advancing our understanding of the impact of sensory feedback on amputees' daily lives. The efficacy and practicality of systems for daily use depend not only on the stimulation sites but also on how the sensory information is represented, as well as individual requirements and preferences [10]. Thus, a flexible system design is required for comparative testing and adoption of haptic feedback in clinical and commercial settings.

Furthermore, the current systems present challenges in terms of reproducibility, primarily stemming from their bulky designs which often feature wired connections and is unsuitable for testing or use outside laboratory [11]. Most designs also lack whole-hand tactile sensor coverage [7, 11, 12], limiting researchers' ability to effectively identify the most essential parts of the prosthetic hand for task execution and the optimal mapping of crucial sensory information to vibration motors. These limitations of the system hinder its applicability, accessibility, and scalability required to conduct comprehensive studies across diverse user profiles and amputation levels.

To address these issues, we propose a wireless reconfigurable haptic feedback system with high-density sensors across the hand to enable a comparative study of different combinations of vibrotactile feedback modalities and stimulation locations. With its affordability and simplicity, this novel design can be reproduced for a wider population to achieve a broader and more inclusive study of the optimal vibrotactile feedback delivery modalities.

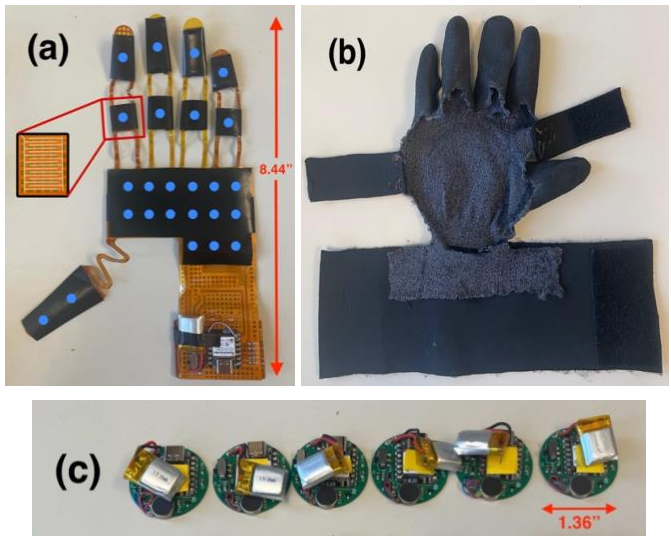

Fig. 1. (a) Tactile sensor fPCB. Blue circles indicate the location of 25 sensors. A picture of the intersection of row and column traces beneath the Velostat™ is shown in red. (b) SensoPatch glove (c) 6 Vibration patches

## II. DESIGN

### A. SensoPatch Feedback System Design

The architecture of the SensoPatch feedback system consists of three main components: a tactile sensor array that collects and transmits pressure information, a set of haptic feedback sticker patches that convert the sensor data into vibration for user, and a glove.

Fig. 1(a) presents an overview of the tactile sensor, which features a flexible printed circuit boards (fPCB), designed via KiCad schematics, for scalable production. The fPCB consists of 5 conductive column traces and 5 conductive row traces that intertwine at key sensory locations in the fingers and palm, forming a 5x5 sensor array with 25 tactile sensors. Velostat™, a piezoresistive material, is placed on the sensor. When the fPCB is pressed, the pressure at each intersection translates to voltage readings, which can be measured using a time-division multiple access (TDMA) approach. To facilitate flexible movement of the thumb, we incorporate a zig-zag trace design to allow the fPCB to stretch. This tactile sensor design is created to fit the TASKA HandGen2.

Fig. 1(b) presents an open back glove with adjustable wrist straps for retrofittability, enabling integration of the tactile sensor onto the prosthetic hand.

Fig. 1(c) presents an overview of the sticker patches, each featuring a vibration motor on a 1.36" diameter circular rigid PCB. The patch can be placed anywhere on the body using Vapon® medical double-sided tape. The modularity of each feedback motor allows users to easily alter the number of motors and customize motor placement.

Both sensing glove and vibration stickers feature Seeed Studio's XIAO BLE nRF52840 for voltage readings, data processing, motor control, and wireless communication. The XIAO is a small, low-cost, and powerful programmable microcontroller which offers Bluetooth Low Energy (BLE) connectivity and built-in battery charging circuit. Additionally, both components include a 3.7 V 100 mAh lithium-ion battery and an on/off slide switch.

SensoPatch is an open-source tool that utilizes off-the-shelf components, which ensures that the device is both cost-effective and easy to assemble. With its accessbility and reproducibility, SensoPatch enables researchers to conduct extensive studies on haptic feedback delivery, ultimately advancing the research toward practical, optimized solutions for amputees. The design files can be found at the following link: https://github.com/angyanisa/SensoPatch

### B. Feedback Modalities

SensoPatch provides a tool to explore the optimal feedback strategies for representing key sensory information, particularly in case of high-density sensors where the number of sensors exceeds the number of vibration motors. To demonstrate this application, we programmed 6 compression modes, which are different methods of compressing tactile data by mapping sensors from various areas of the hand to different numbers of vibration motors, as illustrated in Table I. To convey pressure information, we implemented a binary feedback control, where sensor readings above a threshold trigger vibration motors to produce the same vibration intensity.

TABLE I. COMPRESSION MODES

|  | Number of Motors | | |
| --- | --- | --- | --- |
|  | *1* | *3* | *6* |
| Finger-focused | [hand image] | [hand image] | [hand image] |
| Palm-focused | [hand image] | [hand image] | [hand image] |

A table illustrating the mapping of sensors to vibration motors. Black color indicates area with no feedback. The rest of the colors indicate mapping to different motors. The pressure from all sensors in the area with similar color was averaged and mapped to the same motor. The numbers indicate the order that vibration motors were aligned on the body.

## III. EXPERIMENTAL METHODS

To demonstrate the potential of SensoPatch, we conducted two preliminary experiments focused on optimizing feedback.

### A. Haptic Vibration Discrimination

We performed discrimination experiment to identify the location that offers the best vibration perception. Fig. 2 shows three body parts chosen as an example for investigation: upper arm, shoulder, and lower back. Despite the lack of research on the optimal placement of motors, forearm and upper arm are the most common spots to install haptic device [4]. Thus, one of the locations is included for comparison. The rest of the locations were selected based on criteria of good vibrotactile sensitivity [13], adequate area to accommodate up to 6 vibration motors, and minimal movement, as both passive and active movement can decrease haptic perceptions [14]. In this experiment, approved by the Johns Hopkins Medicine

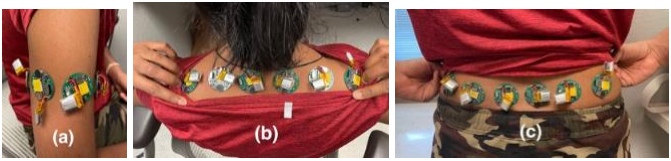

Fig. 2. Motor placement on the (a) upper arm (b) shoulder (c) lower back

Institutional Review Board (JHM IRB), 10 able-bodied participants were recruited. Participants were asked to wear a noise-canceling headphones while the experimenter provided vibration feedback for a duration of 1000 ms. The experiment is divided into 2 parts, (a) and (b).

*a) Vibration Intensity Discrimination:* We assessed the accuracy of vibration intensity perception. During the 5 minute training phase, participants were presented with 3 different vibration levels from one motor and asked to memorize the intensity. During the experiment, participants were instructed to identify the intensity level. Each level was tested 10 times in random order for a total of 30 trials.

*b) Vibration Location Discrimination:* We assessed the ability to distinguish between 6 vibration motors. The motors were equally distributed to maximize the distances at each body location to suit participants' morphology. During the 5 minute training phase, each of the 6 motors was activated sequentially to allow participants to familiarize themselves with the position of vibrations. During the experiment, the stimulation was sent to one of the motors and the participants were asked which motor they believe had been activated. Each motor was tested 5 times in random order for a total of 30 trials. To assess the accuracy of distinguishing feedback from multiple sensors in the case of multi-touch sensations, the training and testing were repeated, but with a stimulation to two motors out of six motors simultaneously. Each combination pair was tested 3 times for a total of 45 trials.

Both sessions were repeated for 3 body locations. All trials at one location were tested together due to the impracticality of switching locations between trials. However, the order of locations tested was randomized for each subject.

### B. Object Manipulation

We assessed the participants' ability to pick up 5 common objects (ball, book, glass, water bottle, and teddy bear), shown in Fig. 3(a), while being blindfolded and relying on haptic feedback using the best location previously found to evaluate the effect of compression modes on object manipulation. Three able-bodied participants were equipped with a bypass prosthetic and a Myo™ armband to control the TASKA HandGen2 myoelectric hand via EMG (Fig. 3(c)). The subjects were trained to open and close the prosthetic hand. Before the trials for each compression mode, subjects familiarized themselves with feedback from grabbing different objects. During the experiment, each object was presented 3 times in random order. The average time taken to pick up each object was recorded. The experiment was repeated for 6 compression modes, and the order was randomized among participants. An example of the tactile data obtained when a participant grabbed a ball is shown in Fig. 3(d).

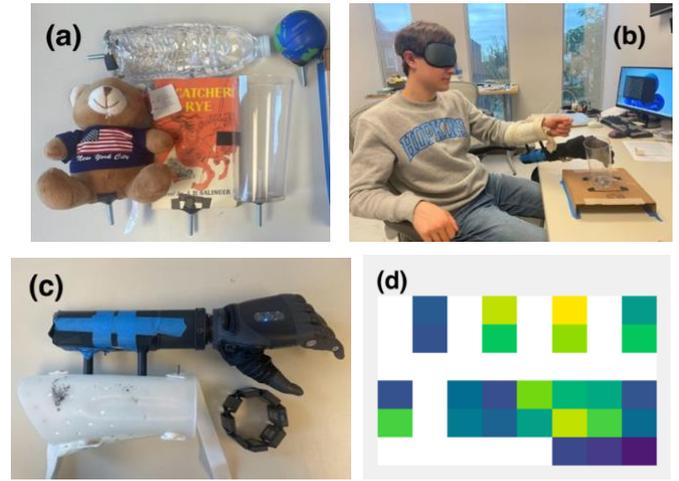

Fig. 3. (a) Five objects were selected as representative of the variation in stiffness, and shape that might be encountered in daily living. (b) Object manipulation task (c) Bypass prosthetic (d) Visualization of the pressure distribution on all fingers (from left to right: thumb to pinky) and palm when the subject grabbed a ball. Lighter color represents higher pressure.

### IV. RESULTS

#### A. Haptic Vibration Discrimination Results

Fig. 4(a)-(c) display the true versus predicted intensity levels across all 10 participants. The accuracy of the upper arm

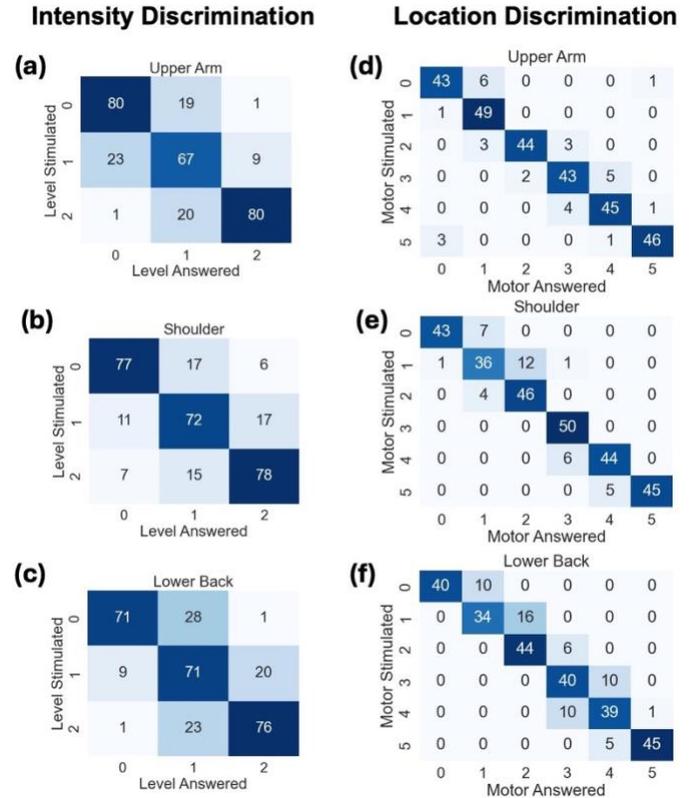

Fig. 4. (a)-(c) Vibration intensity discrimination results on upper arm, shoulder, and lower back respectively (d)-(f) Single-point vibration location discrimination results on upper arm, shoulder, and lower back respectively.

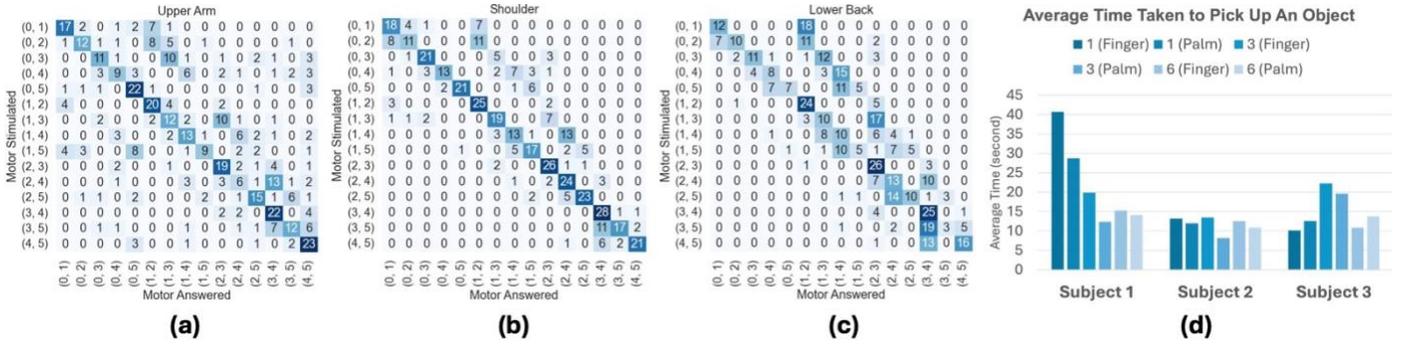

Fig. 5. (a)-(c) Two-touch vibration location discrimination results on upper arm, shoulder, and lower back respectively (d) Bar graph depicting the average time taken to pick up an object for all 3 subjects. For each subject, the results from 6 compression modes are shown. The data were normalized to the average across all objects and all trials for each condition. Less time indicates that the compression mode for the feedback effectively helps subjects pick up objects.

and the shoulder were tied at 76%, followed by the accuracy of the lower back at 73%. Similarly, the confusion matrices from the location discrimination tests are shown in Fig. 4(d)-(f) and 5(a)-(c). The result from a single motor discrimination indicates that the best score was obtained on the upper arm (90%), followed closely by the shoulder (88%) and lower back (81%). However, the highest accuracy of identifying two motors simultaneously obtained on the shoulder (66%) was substantially higher than the accuracy on the upper arm (49%) and lower back (42%). Overall, the shoulder offered the most reliable feedback perception (77%) considering both intensity and location of vibrations, compared to the upper arm (72%) and lower back (65%). Thus, the shoulder was used in the object manipulation task.

*B. Object Manipulation Results*

The average time taken to pick up an object for each compression mode is shown in Fig. 5(d). There was no observable trend. Using Friedman's ANOVA test, the p-value was above the 0.05 threshold, indicating that the effect of compression modes cannot be concluded at this sample size.

## V. DISCUSSION AND CONCLUSION

The preliminary experiments show the potential of SensoPatch as a research tool for optimizing various aspects of vibrotactile feedback. Specifically, the wireless and customizable feature of SensoPatch is designed for experimentation on the location for feedback delivery. With its modularity, SensoPatch is designed to improve feedback perception since motors with larger spacings can be better discriminated [15]. Additionally, the high-density sensor array not only enables users to gain sensory information from most area on the prosthetic hand but also serves as an instrument for the investigation of optimal compression methods.

The example data from the vibration discrimination experiment suggested that several factors could affect the performance on different body locations. Although we observed a comparable performance between feedback perception on the upper arm and on the shoulder for the intensity and single motor discrimination tasks, the shoulder showed greater accuracy in distinguishing between two vibration motors (Fig. 5(a)-(c)). This could be because the first and last motor might serve as reference points for determining the positions of all motors. Due to the circular arrangement on the upper arm where the first and last motor are adjacent, confusion about the remaining positions may lead to more scattered misidentifications. Further studies on amputees are needed to identify the key factors influencing the ability to discriminate vibration feedback. Nevertheless, these findings suggested that the most commonly used haptic device designs, such as armbands, may not be the most effective for feedback delivery. SensoPatch enables future comparative studies that are necessary for exploring the optimal placement of feedback stimulators before haptic feedback system can be integrated into daily use.

The results from the object manipulation task in Fig. 5(d) show that none of the conditions had a statistically significant effect on the task performance, suggesting that more participants may be needed to investigate the relationship. Moreover, the result of this experiment might be influenced by learning over time, leading to an improved performance over successive trials. Future work should consider changing the location of the objects between trials to prevent participants from memorizing positions instead of relying on haptic feedback to complete the task. The experiment should also be repeated for more trials to address potential outliers. Additionally, the variations observed between subjects in this experiment may suggest that individuals have personal preferences for different feedback modalities.

This work details the design and provides original files for recreation of SensoPatch as well as examples to demonstrate the potential of SensoPatch using three stimulation locations and six compression modes. Future researchers can leverage the reproducibility and unique adaptability of SensoPatch to conduct experiment using other motor placement, including a combination of multiple body locations. In addition, other feedback parameters can be explored such as proportional feedback control, derivative feedback control, other number of vibration motors, other vibration patterns, or different sensory information conveyed.

ACKNOWLEDGMENT

We thank the participants for their involvement in the study and Keqin Ding for her guidance throughout the experiments.

## REFERENCES

[1] S. Raspopovic, G. Valle, and F. M. Petrini, "Sensory feedback for limb prostheses in amputees," Nature Materials, vol. 20, no. 7. Springer Science and Business Media LLC, pp. 925–939, Apr. 15, 2021. doi: 10.1038/s41563-021-00966-9.

[2] A. W. Shehata, M. Rehani, Z. E. Jassat, and J. S. Hebert, "Mechanotactile sensory feedback improves embodiment of a prosthetic hand during active use," Frontiers in Neuroscience, vol. 14. Frontiers Media SA, Mar. 26, 2020. doi: 10.3389/fnins.2020.00263.

[3] M. B. S. Gallone, "Development of a wearable haptic feedback device for upper limb prosthetics through sensory substitution," Electronic Thesis and Dissertation Repository, 2021. Accession number: 7873

[4] M. Nemah, C. Low, O. Aldulaymi, P. Ong, A. Ismail, and A. Qasim, "A review of non-invasive haptic feedback stimulation techniques for upper extremity prostheses," International Journal of Integrated Engineering, vol. 11, pp. 030, 2019. doi: 10.30880/ijie.2019.11.01.030.

[5] M. Nabeel, K. Aqeel, M. N. Ashraf, M. Awan, and M. Khurram, "Vibrotactile stimulation for 3D printed prosthetic hand," Proceedings of the IEEE International Conference on Robotics and Artificial Intelligence (ICRAI), pp. 202-207, 2016. doi: 10.1109/ICRAI.2016.7791254.

[6] T. Li, H. Huang, C. Antfolk, J. Justiz, and V. Koch, "Tactile display on the remaining hand for unilateral hand amputees," Current Directions in Biomedical Engineering, vol. 2, 2016. doi: 10.1515/cdbme-2016-0089.

[7] A. Aug, A. Slepyan, E. Levenshus and N. Thakor, "Haptic Touch: A retrofittable tactile sensing glove and haptic feedback armband for scalable and robust sensory feedback," 2022 9th IEEE RAS/EMBS International Conference for Biomedical Robotics and Biomechatronics (BioRob), Seoul, Korea, Republic of, 2022, pp. 01-06, doi: 10.1109/BioRob52689.2022.9925475.

[8] E. Sariyildiz, F. Hanss, H. Zhou, M. Sreenivasa, L. Armitage, R. Mutlu, and G. Alici, "Experimental evaluation of a hybrid sensory feedback system for haptic and kinaesthetic perception in hand prostheses," Sensors (Basel), vol. 23, no. 20, pp. 8492, Oct. 16, 2023. doi: 10.3390/s23208492.

[9] B. Peerdeman et al., ''Myoelectric forearm prostheses: State of the art from a user-centered perspective,'' J. Rehabil. Res. Develop., vol. 48, no. 6, pp. 719–738, 2011.

[10] B. Stephens-Fripp, G. Alici and R. Mutlu, "A review of non-invasive sensory feedback methods for transradial prosthetic hands," IEEE Access, vol. 6, pp. 6878-6899, 2018, doi: 10.1109/ACCESS.2018.2791583.

[11] P. B. Shull and D. D. Damian, "Haptic wearables as sensory replacement, sensory augmentation and trainer – a review," Journal of NeuroEngineering and Rehabilitation, vol. 12, no. 1. Springer Science and Business Media LLC, Jul. 20, 2015. doi: 10.1186/s12984-015-0055-z.

[12] M. Markovic et al., "The clinical relevance of advanced artificial feedback in the control of a multi-functional myoelectric prosthesis," Journal of NeuroEngineering and Rehabilitation, vol. 15, no. 1. Springer Science and Business Media LLC, Mar. 27, 2018. doi: 10.1186/s12984-018-0371-1.

[13] A. Wilska, "On the vibrational sensitivity in different regions of the body surface," Acta Physiologica Scandinavica, vol. 31, pp. 285-289, 1954. doi: 10.1111/j.1748-1716.1954.tb01139.x.

[14] I. Karuei, K. MacLean, Z. Foley-Fisher, R. MacKenzie, S. Koch, and M. El-Zohairy, "Detecting vibrations across the body in mobile contexts," in Proceedings of the SIGCHI Conference on Human Factors in Computing Systems (CHI '11). Association for Computing Machinery, New York, NY, USA, pp. 3267–3276, 2011. doi: 10.1145/1978942.1979426.

[15] M. Guemann, S. Bouvier, C. Halgand et al., "Effect of vibration characteristics and vibror arrangement on the tactile perception of the upper arm in healthy subjects and upper limb amputees," Journal of NeuroEngineering and Rehabilitation, vol. 16, no. 1, p. 138, 2019. doi: 10.1186/s12984-019-0597-6.